\documentclass[epjCONF, onecolumn]{svjour}
\usepackage{graphics}
\usepackage[varg]{txfonts} 
\usepackage[latin1]{inputenc}
\session-title{Assembling the Puzzle of the Milky Way}
\begin{document}
\title{Stellar kinematical signatures of disk non-axisymmetries in the extended solar neighbourhood}
\author{Benoit Famaey\inst{1,2}\fnmsep\thanks{\email{benoit.famaey@astro.unistra.fr}} \and Arnaud Siebert\inst{1} \and Ivan Minchev\inst{3}, for the RAVE collaboration}
\institute{Observatoire Astronomique de Strasbourg, CNRS UMR 7550, France 
\and Argelander Institute for Astronomy, Bonn, Germany 
\and Leibniz-Institut f\"ur Astrophysik Potsdam (AIP), Germany}
\abstract{Bars and spirals are among the main drivers of the secular evolution of galactic disks, and it is, thus, of prime importance to better understand their exact nature and dynamics. In this respect, the Milky Way provides a unique laboratory in which to study a snapshot of their detailed dynamical influence in six-dimensional phase-space. With the advent of present and future astrometric and spectroscopic surveys, such six-dimensional phase-space data can be obtained for stars in an increasingly large volume around the Sun. Here, we review the signatures of -- and constraints on -- disk non-axisymmetries obtained from recent stellar kinematical data sets, including (i) the detection of resonant moving groups in the solar neighbourhood, (ii) the non-zero value of the Oort constants $C$ and $K$, and (iii) the detection of a radial velocity gradient of $\sim 4\,$km/s/kpc in the extended solar neighbourhood ($d<2$~kpc). 
} 
\maketitle
\section{Introduction}
\label{intro}

Over the last decade, large progress has been made in understanding galactic disks formation and evolution in a cosmological context \cite{BinneyT}, despite some remaining discrepancies between simulations and observations. It has for instance become clear that the hierarchical merging history of galaxies may not necessarily be the dominant process in the formation and evolution of disks, but that internal secular evolution might also play a major role. Among the main drivers of this secular evolution of galactic disks are their instabilities and associated non-axisymmetric perturbations, including the bar and spiral arms. Questions about their nature -- transient, quasi-stationary, or both types co-existing -- , about their detailed structure and dynamics -- e.g., amplitude and pattern speed -- , as well as questions about their influence on secular processes such as stellar migration, are all essential elements for a better understanding of galactic evolution. The Milky Way provides a unique laboratory in which a snapshot of the dynamical effect of present-day disk non-axisymmetries can be studied in great details, in order to help answering the above questions.

Most of  our knowledge of the structure and dynamics of the bar and spirals of the Milky Way presently comes from gas \cite{Bissantz,combes,Ma}, and notably from its observed longitude-velocity ($l-v$) diagram. However, with the advent of present and future spectroscopic and astrometric surveys, six-dimensional phase-space information for {\it stars} in an increasingly large volume around the Sun will allow us to set new dynamical constraints on the non-axisymmetric perturbations of the Galactic potential, thereby better constraining their nature and influence on the secular evolution of the Galactic disk. We summarize hereafter the present-day signatures and constraints currently obtained from stellar kinematics. This includes the following list:
\begin{itemize}
\item[{\bf 1.}] Kinematics of nearby stars w.r.t. the Local Standard of Rest (LSR)
\subitem {\bf 1. i.} Moving groups with a resonant origin (see Sect.~2)
\subitem {\bf 1. ii.} Oort constants $C$ and $K$ (see Sect.~3)
\item[{\bf 2.}]  Kinematics of more distant stars
\subitem {\bf 2. i.} Net outwards or inwards flow at any given position (see Sect.~4)
\subitem {\bf 2. ii.} Velocity distribution w.r.t. the LSR differing at $l$ and $-l$
\item[{\bf 3.}] Kinematics of the LSR itself (may contain a Galactocentric radial component, see Sect.~4)
\end{itemize}

\section{Moving groups}
\label{sec:1}

\begin{figure}
\centering
\resizebox{0.3\columnwidth}{!}{%
\includegraphics{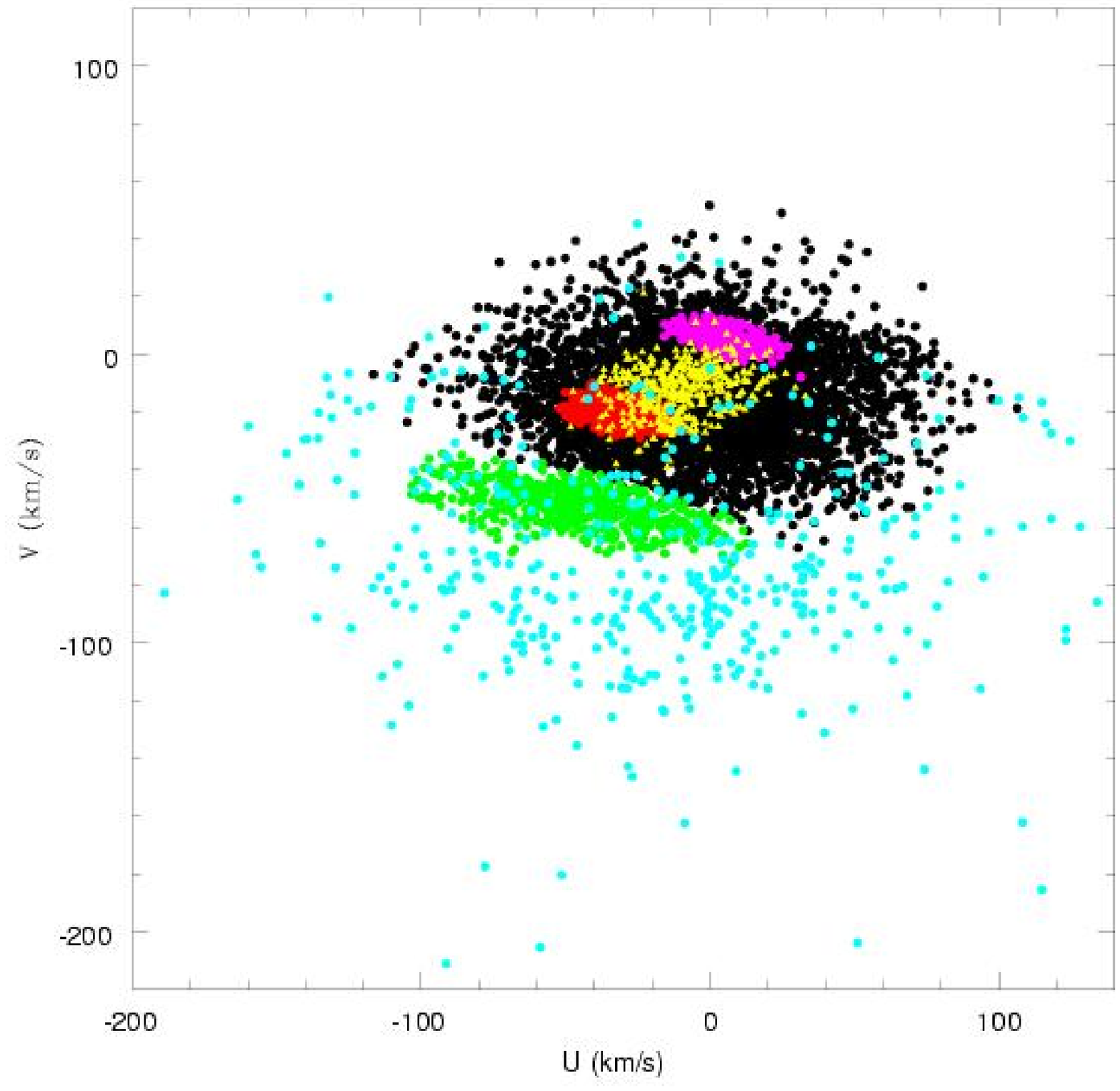}}
\resizebox{0.25\columnwidth}{!}{%
\includegraphics{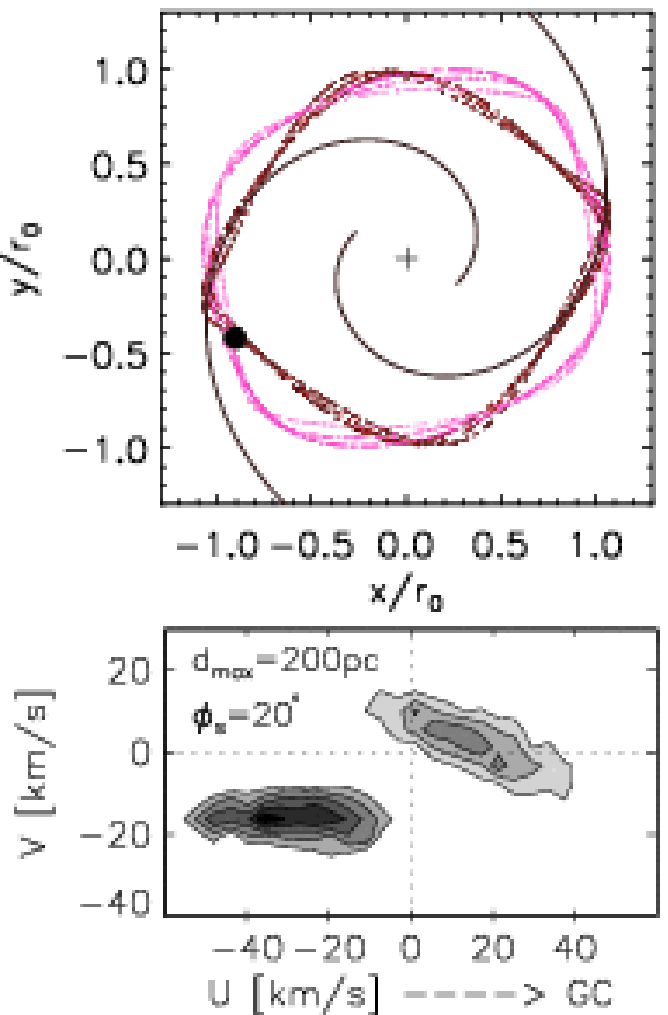}}
\caption{Left panel: 6691 local Hipparcos-CORAVEL giant stars \cite{f05} plotted in the $UV$-plane ($U$ being the velocity towards the Galactic center,
 $V$ the velocity in the direction of Galactic rotation, both with respect to
 the Sun). The distribution is clumpy and six different groups have been identified through a maximum likelihood method. The {\it Hyades} group is in red, {\it Sirius} in magenta, and {\it Hercules} in green \cite{f05}. Top right panel: the effect of a steady-state 2-armed spiral structure with $\Omega_s / \Omega_0= 0.65 $ on orbits near the 4:1 inner resonance \cite{pomp}. For a Sun orientation at $20^\circ$ w.r.t. a concave arm, two families of square-shaped closed orbits enter the solar neighborhood (black filled circle). The galactocentric axes are in units of the galactocentric radius of the Sun $R_0$. Bottom right panel: the effect on the $UV$-plane for the configuration shown in the top-right panel (selecting test particles in a 200~pc circle around the Sun). Each orbital family gives rise to a moving group in velocity space. We can associate the dense clump at $(U,V)\approx(-35,-17)$ km/s with the Hyades (red on the Left Panel) and the shallow one  at $(U,V)\approx(10,0)$ km/s with Sirius (magenta on the Left Panel).}
\label{fig:1}      
\end{figure}

It has been known for  a long  time that the  {\it local} velocity  field in  the solar neighbourhood is  clumpy, and that most  of the observed clumps  are made of spatially unbound groups of stars, called moving groups. While some of the smallest velocity clumps are probably due to the evaporation over time of star clusters \cite{desilva}, modern data have firmly established that the three main low-velocity moving groups, the so-called Hercules, Hyades, and Sirius moving groups (see Fig.~1),  are made of stars of very different ages and chemical compositions, so that the clumping cannot be due to irregular star formation \cite{bovy,De1,f05,f08,pomp}. At radii where the angular speed in the rotating frame of a non-axisymmetric perturber and the radial epicyclic frequency are in $m:1$ resonance (i.e., $m$ epicyclic oscillations in one rotation) , closed orbits are {\it trapped} at resonance and can create precisely this type of kinematic group in local velocity space. However, a plethora of models for the bar and spiral are able to reproduce these {\it local} moving groups, meaning that precise data {\it outside} of the solar neighbourhood will be needed to disentangle the various models (for a review, see \cite{Antoja1,Antoja2,Mo}). There is now a general agreement that Hercules (see Fig.~1) is linked to the bar's outer 2:1 resonance, which has to be located just inside the Solar radius \cite{Dehnen}, in accordance with constraints from gas flows \cite{Bissantz}, meaning that the pattern speed of the bar is $\Omega_b/\Omega_0 \simeq 2$ (where the angular speed at the Sun is $\Omega_0 \simeq 30 \,$km/s/kpc). For the other moving groups, the situation is less clear: transient spirals with corotation or inner 2:1 resonance close to the Sun, as well as recent bar growth with outer 2:1 resonance close to the Sun,  have all been claimed to be responsible for some moving groups \cite{desim,McMil,Minbar,sell}. An interesting possibility, illustrated in Fig.~1,  is that the 4:1 inner resonance of a 2-armed spiral structure splits the velocity distribution into two features corresponding to two square-shaped orbital families. One of them is consistent with the Hyades, the other with Sirius \cite{pomp,minquil}. With this explanation, the Sun should thus be at the same time close to the 2:1 outer resonance of the bar ($\Omega_b/\Omega_0 \simeq 2$) and to the 4:1 inner resonance of the 2-armed outer spiral pattern ($\Omega_s / \Omega_0 \simeq 0.6$). Note that since resonance overlap results in chaotic behavior, we expect that such a resonance overlap could give rise to strong stochastic radial migration of stars inside the disk \cite{Minfam1,Minfam2}.

\section{Oort constants}
\label{sec:2}

The effects of non-axisymmetric perturbations in the Galactic plane can also be analyzed locally by Taylor expanding to first order the planar velocity field in the LSR cartesian frame (an approximation roughly valid up to a distance of 2~kpc). This is done by generalizing the classical Oort constants to the case of a non-axisymmetric disk. The cold velocity field generated by closed orbits with respect to the LSR velocity can be expressed as:
\begin{equation}
v_i = H_{ij} x_j + {\cal O} (x^2)
\end{equation}
where
\begin{equation} \label{eq:oort-def-b}
  \mathbf{H} =  \left(
    \begin{array}{c@{\;\;}c}
      K+C & A-B \\
      A+B & K-C
    \end{array}
  \right),
\end{equation}
and the parameters $A$, $B$, $C$, and $K$ are called the Oort constants. They measure the local divergence ($K$), vorticity ($B$), azimuthal shear ($A$) and radial shear ($C$) of the velocity field generated by closed orbits. Axisymmetry implies $C=K=0$ (but not the reverse, they can, e.g., be zero if the main non-axisymmetric perturbation is symmetric w.r.t. the Sun-Galactic center axis). Proper motions of a large sample of stars allow for a measurement of $A$, $B$, and $C$, while line-of-sight velocities projected onto the Galactic plane give access to $A$, $C$ and $K$. While old, rather imprecise, data were actually compatible with the axisymmetric values $C=K=0$ \cite{Kuijken94}, a modern analysis of ACT/Tycho-2 proper motions, after corrections for the mode-mixing and asymmetric drift, yielded $C = -10 \,$km/s/kpc for the red giants population \cite{Olling}(with a typical $\sigma_R \sim 40 \,$km/s). What was even more surprising is that $C$ was found to vary approximately linearly with color (and thus mean age) along the main sequence, from $C = 0$ for blue (young, and low velocity dispersion) samples, to $C = -10$ for late-type stars (old, and high-dispersion), contrary to the naive expectation that non-axisymmetric structure would mainly affect the low-dispersion populations. Numerical integration of test-particles in a gravitational potential including the Galactic bar and no spiral arms \cite{MinNor} has shown that this effect of increasing $|C|$ with velocity dispersion could be qualitatively reproduced if the Sun is located outside the outer 2:1 resonance of the bar with the right phase (see Fig.~2), in accordance with the position of the local Hercules moving group and with gas flows. This is because high-velocity dispersion stars do not probe the same guiding radii as low velocity dispersion ones, and tend to shift the ``effective resonance" radius radially outwards. However, to quantitatively reproduce the observed value of $C$, one would probably need a combination of both the bar and spiral.

The measurement of the Oort constant $K$ was, on the other hand, recently performed with the Radial Velocity Experiment (RAVE \cite{DR3}) in the longitude interval $-140^\circ < l < 10^\circ$, thanks to the line-of-sight velocities (projected onto the Galactic plane) of 213713 stars (dominated by red giants) with spectro-photometric distances $<2 \,$kpc from the Sun. This analysis \cite{flow} confirmed the above proper-motion value of $C=-10$, and found a value $K= +6$, also different from zero, and from the one found for very young stars \cite{Fernandez}. This value would actually imply a Galactocentric radial velocity gradient of $C+K = \partial V_R / \partial R \simeq - 4\,$km/s/kpc in the extended solar neighbourhood (see Sect.~4).

\begin{figure}
\centering
\resizebox{0.45\columnwidth}{!}{%
\includegraphics{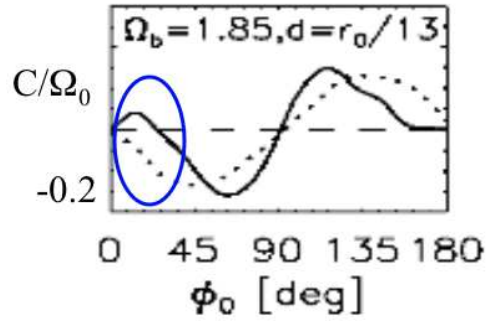}}
\caption{The value of the Oort constant $C$ (in units of $\Omega_0$) as a function of the bar angle, as measured from a simulated sample of cold orbits (solid, $\sigma_R \sim 10 \,$km/s) and hotter ones (dotted, $\sigma_R \sim 40 \,$km/s) in a Milky Way-like potential with a central bar \cite{MinNor}. In the encircled region, one remarkably has $|C|_{\rm hot} > |C|_{\rm cold}$.}
\label{fig:2}      
\end{figure}

\section{Radial velocity gradient}
\label{sec:3}

To check the actual existence of the radial velocity gradient implied by the above-measured value of $C+K$, we \cite{flow} looked at the projection onto the plane of the mean line-of-sight velocity as  a function of $d \, {\rm cos} l \, {\rm cos} b$ for $|l|<5^\circ$, both for the full RAVE sample and for red clump candidates (with an independent method of distance estimation).  We  then compared the  observed mean velocities to  the expected
velocities for  a thin disk  in circular rotation with an additional radial gradient. 
We clearly confirmed that RAVE data  are  not compatible  with  a disk  in circular  rotation because the  mean projected  line-of-sight velocity is  systematically lower than the expected mean velocity. They are, on the other hand, roughly compatible with a linear gradient of  $\partial V_R / \partial R \simeq - 4\,$km/s/kpc.

Making use of the full information available, including PPMX proper motions, we \cite{flow} also computed the full  2D velocity field. Converting the $(U,V)$ velocities in the Heliocentric reference frame into the $(V_R,V_\phi)$ Galactocentric coordinates requires knowing the Galactocentric radius of the Sun $R_0$, the Sun's peculiar velocity with respect to the LSR, and the motion of the LSR with respect to the Galactic center:
\begin{eqnarray}
V_R = - (U + U_\odot + U_{LSR}) \, {\rm cos} \alpha + (V + V_\odot + V_{LSR}) \, {\rm sin} \alpha  \,\, , \nonumber\\
V_\phi = + (U + U_\odot + U_{LSR}) \, {\rm sin} \alpha + (V + V_\odot + V_{LSR}) \, {\rm cos} \alpha\,\, ,
\end{eqnarray}
where $\alpha$ is the Star-Galactic center-Sun angle. We first assume $R_0=7.8 \,$kpc for the Sun's radius, $U_{LSR}=0 \,$km/s, $V_{LSR}=247 \,$km/s, and $(U_\odot,V_\odot)=(11,12)\,$km/s \cite{Schonrich}: the result for the map of $\langle V_R \rangle$ is plotted on Fig.~3.

There are four interesting features about this result:
\begin{itemize}
\item[{\bf 1.}] The gradient in $\langle V_R \rangle$ is not really linear. It is almost flat at distances $d< 1 \,$kpc, and becomes steep at larger distances from the Sun in the direction of the Galactic center. The gradient is not only a function of distance from the Sun but also of longitude. 
\item[{\bf 2.}] The gradient affects a sample dominated at large distances by red giants, with a typical velocity dispersion $\sigma_R \sim 40 \,$km/s. This is in line with the finding that the Oort constant $C$ is nonzero for such high-$\sigma$ samples. This might indicate that the Galactic bar plays a role in shaping the gradient (see Sect.~3).
\item[{\bf 3.}] The gradient affects stars substantially above the plane, keeping in mind that RAVE lines of sight are typically at $b >\sim 20^\circ$. The zone where the gradient is steep is populated with stars with typically $|z| \sim 500 \,$pc.
\item[{\bf 4.}] Assuming $U_{LSR}=0$ puts the Sun at a {\it privileged} position where $\langle V_R \rangle =0$. 
\end{itemize}

Concerning the last point, independently measuring $U_{LSR}$, or simply measuring the Sun's velocity w.r.t. the Galactic center, would need either (i) measuring the mean velocity of gas clouds along the Sun-Galactic center line-of-sight with HI absorption features in the radio emission of Sgr A*, or (ii) using combined spectroscopy and astrometry of S2 stars around Sgr A*, or (iii) measuring $\langle V_R \rangle (\alpha)$ for  a {\it well-mixed} stellar halo sample (if that physically exists), with full sky coverage\footnote{Note that the answer also depends on how exactly one defines the Galactic center, as it is not obvious that the position of the central supermassive black hole would necessarily coincide with the center of mass of the dark halo.}. Results from the first method indicate $U_{LSR} = 0 \pm 0.25 \,$km/s \cite{Radha}, results from the second one indicate $U_{LSR} = 20 \pm 33 \,$km/s \cite{Ghez}, and results from the third one indicate either  $U_{LSR} = -2.5 \pm 2.2 \,$km/s \cite{Gould} or  $U_{LSR} = 9 \pm 2.5 \,$km/s \cite{Smith}. Given these large uncertainties, we plot in Fig.~4 the 2D $\langle V_R \rangle$ field assuming, as an example, $U_{LSR} = 7 \,$km/s and $U_{LSR} = -9 \,$km/s. The first case is interesting because the outwards flow then does not affect anymore stars located substantially above the plane. As a matter of fact, $\langle V_R \rangle$ would then be negative in the Sun's neighbourhood and reach $0$ at $d \sim 1.5 \,$kpc and $z \sim 500 \,$pc in the RAVE sample. Clearly, until one has an independent and precise confirmation of the fact that $\langle V_R \rangle =0$ at the Sun's position, the value of $U_{LSR}$, or equivalently the Galactocentric Sun's radial velocity $V_{R\odot}$ should be kept as a free parameter in fitting models to kinematic data.

\begin{figure}
\centering
\resizebox{0.5\columnwidth}{!}{%
\includegraphics{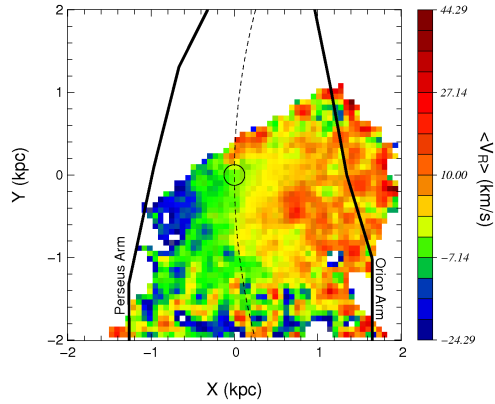}}
\caption{Mean Galactocentric radial velocity $\langle V_R \rangle$ from RAVE, as a function of position in the extended solar neighbourhood \cite{flow}. X increases in the direction of the Galactic center, Y is positive towards the Galactic rotation. The  locations of the  nearest spiral arms are  indicated.  The open circle  delimitates a sphere 125~pc in radius  around the Sun. Stars tend to move on average towards the {\it outer} Galaxy as one leaves the solar neighbourhood in the direction of the Galactic center. Note that $\langle V_R \rangle =0$ at the Sun's position {\it by construction}, because one assumes $U_{LSR}=0$. One sees that the value of $\langle V_R \rangle$ is not only a function of distance from the Sun but also of longitude.}
\label{fig:3}      
\end{figure}

\begin{figure}
\centering
\resizebox{0.45\columnwidth}{!}{%
\includegraphics{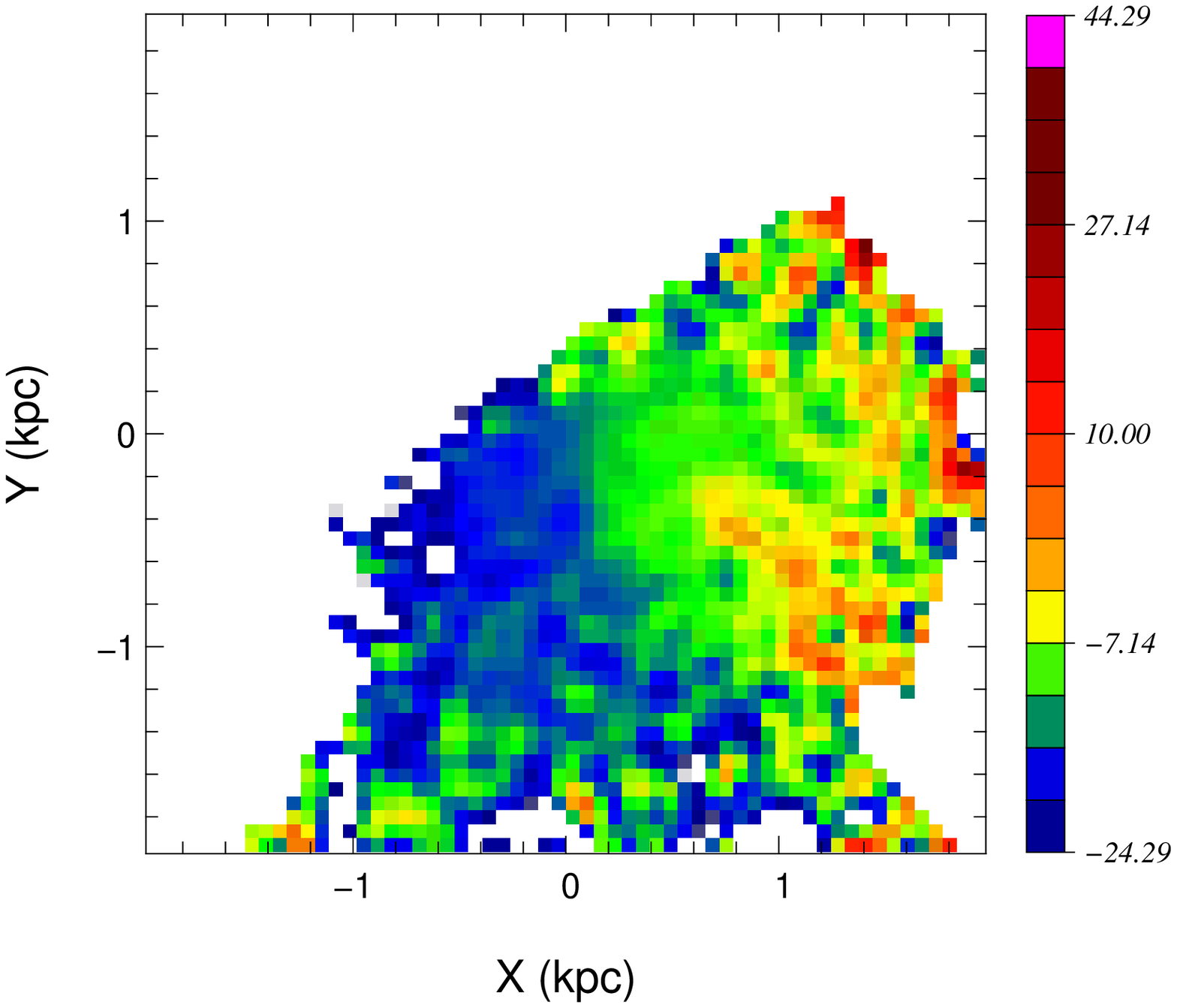}}
\resizebox{0.45\columnwidth}{!}{%
\includegraphics{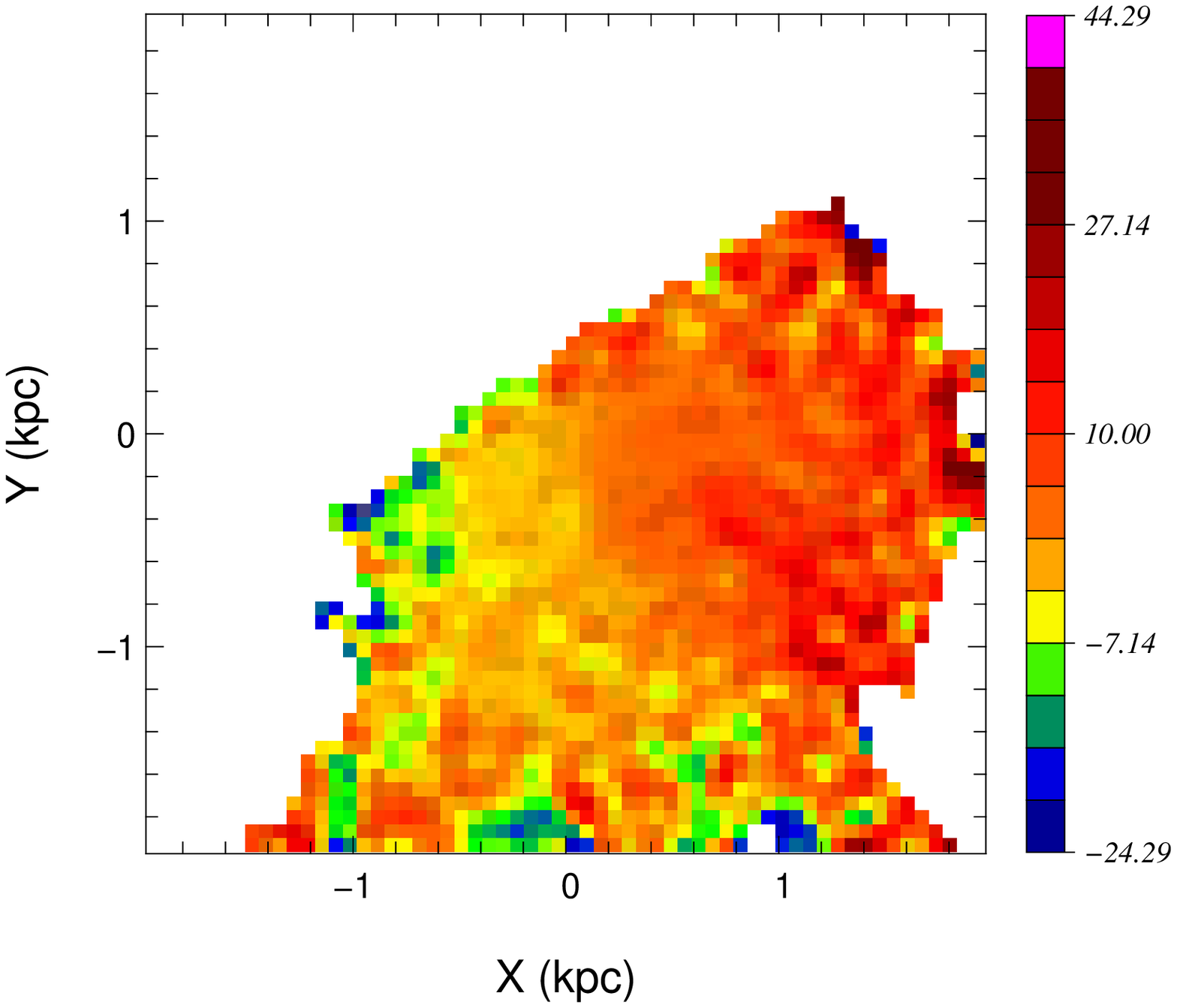}}
\caption{Left Panel: Same as Fig.~3 but assuming $U_{LSR} = +7 \,$km/s, i.e. a mean motion of local stars towards the Galactic center (or, equivalently, a Galactocentric Sun's velocity $V_{R\odot}= -18 \,$km/s). Note that $\langle V_R \rangle$  reaches $0$ at a large distance and thus large height $z$. Right Panel:  Same as Fig.~3 but assuming $U_{LSR} = -9 \,$km/s, i.e. a mean motion of local stars towards the anticenter (or, equivalently, a Galactocentric Sun's velocity $V_{R\odot}= -2 \,$km/s).} 
\label{fig:4}      
\end{figure}

\section{Conclusion}
\label{sec:4}
We have briefly reviewed the current signatures of disk non-axisymmetries obtained from recent stellar kinematical data. The measurement of Oort constants and the detection of moving groups in the solar neighbourhood bring interesting constraints on the Milky Way bar and spiral, but the fit is not unique. It is, thus, of prime importance to have access to precise six-dimensional phase-space data for large samples of stars at different positions inside the Galactic disk in order to discriminate between the various models. The recent detection of a radial velocity gradient of $\sim 4\,$km/s/kpc in the extended solar neighbourhood brings a new important constraint to these models, but it will need independent confirmation at lower latitudes, as well as a determination of how it varies with the dispersion of subpopulations. It will also be mandatory to investigate the possible existence of systematic distance errors, and whether these can, at least partially, lead to the observed effect. Then, short-term perspectives will include the modelling of such a gradient with 2D barred models and with spiral models, before combining both bar and spiral modes in 3D self-consistent models. Such models should a priori keep the Galactocentric Sun's velocity as a free parameter when fitting data. Finally, these models will ultimately have to include hydrodynamics to check whether stellar kinematical constraints are in accordance with gaseous ones. This whole enterprise will hopefully help gaining a better understanding of the present and past structure of disk non-axisymmetries in the Galaxy, and of their influence on the disk evolution.

\end{document}